\pdfoutput=1
\documentclass[11pt]{article}

\usepackage{ACL2023}

\usepackage{times}
\usepackage{latexsym}
\usepackage{amsmath}
\usepackage{amssymb}
\usepackage{algorithm}
\usepackage{algpseudocode}
\usepackage{indentfirst}

\usepackage{enumitem}

\usepackage[T1]{fontenc}
\usepackage[utf8]{inputenc}

\usepackage{microtype} 

\usepackage{inconsolata}
\usepackage{graphicx}

\usepackage{tabularx}

\usepackage{hyperref}
\usepackage{url}
\usepackage{graphicx}
\usepackage{tcolorbox}
\tcbuselibrary{skins}
\usepackage{amsmath}

\usepackage{cleveref}
\Crefrangeformat{table}{Table~#3#1#4--#5#2#6} % 自定义引用格式

\usepackage{listings}
\definecolor{lightgray}{gray}{0.95}
\lstdefinestyle{myprompt}{
    basicstyle=\ttfamily\fontsize{7pt}{8pt}\selectfont,
    frame=none,
    breaklines=true,
    backgroundcolor=\color{lightgray},
    breakatwhitespace=true,
    breakindent=0pt,
    escapeinside={(*@}{@*)},
    numbers=none,
    numbersep=5pt,
    xleftmargin=5pt,
}
\tcbset{
  myaibox/.style={
    top=10pt,
    colback=white,
    colframe=black,
    colbacktitle=black,
    enhanced,
    center,
    attach boxed title to top left={yshift=-0.1in,xshift=0.15in},
    boxed title style={boxrule=0pt,colframe=white,},
  }
}
\newtcolorbox{AIbox}[2][]{myaibox,title=#2,#1}
\usepackage{multirow}

\title{ROGRAG: A Robustly Optimized GraphRAG Framework}
\author{
  Zhefan Wang\textsuperscript{1} \quad
  Huanjun Kong\textsuperscript{1}\footnotemark[2] \quad
  Jie Ying\textsuperscript{1} \quad
  Wanli Ouyang\textsuperscript{1,3} \quad
  Nanqing Dong\textsuperscript{1,2,}\footnotemark[1] \quad \\
  \textsuperscript{1}Shanghai Artificial Intelligence Laboratory\textsuperscript{2}Shanghai Innovation Institute \\
  \textsuperscript{3}Chinese University of Hong Kong\\
}

\begin{document}
\maketitle

{
\renewcommand{\thefootnote}{\fnsymbol{footnote}}
\footnotetext[1]{Corresponding author.}
{\fnsymbol{footnote}}
\footnotetext[2]{Project lead.}
}

\begin{abstract}
\setlength{\parindent}{1em}
\hspace*{\parindent}
% Graph-based Retrieval Augmented Generation (GraphRAG) has achieved significant performance gains in question-answering systems. However, its pipeline is highly engineering-oriented, involving numerous components and parameters that require tuning. This makes it difficult to determine whether the performance gains are due to pipeline optimization or internal parameters. Additionally, with Large Language Model (LLM) training tokens approaching their limits, many public Retrieval Augmented Generation (RAG) QA datasets have been incorporated into LLM training sets. The different input prompts of LLMs can affect the generated results, and identifying which key tokens trigger appropriate outcomes is challenging for non-LLM training personnel. This uncertainty decreases the differentiation of RAG results, leading to inaccurate pipeline evaluations (for example, our tests on citation RAG accuracy improvements on certain small LLMs yielded random results). Given these challenges, this paper does not propose a new method, but selects a domain-specific dataset with low LLM scores. By integrating multiple approaches (with over 18k lines of code), we conducted parameter comparisons and ablation studies. Ultimately, we developed a GraphRAG pipeline capable of streaming responses. We release the code\footnote{\url{https://github.com/tpoisonooo/ROGRAG} is our implementation, while the dataset cannot be fully open-sourced due to licensing restrictions}. 
Large language models (LLMs) commonly struggle with specialized or emerging topics which are rarely seen in the training corpus. Graph-based retrieval-augmented generation (GraphRAG) addresses this by structuring domain knowledge as a graph for dynamic retrieval. However, existing pipelines involve complex engineering workflows, making it difficult to isolate the impact of individual components. It is also challenging to evaluate the retrieval effectiveness due to the overlap between the pretraining 
 and evaluation datasets. 
In this work, we introduce \textbf{ROGRAG}, a \textbf{R}obustly \textbf{O}ptimized \textbf{G}raph\textbf{RAG} framework. Specifically, we propose a multi-stage retrieval mechanism that integrates dual-level with logic form retrieval methods to improve retrieval robustness without increasing computational cost. 
% Furthermore, we implement various result verification methods and conduct extensive ablation experiments to evaluate the effectiveness of each component.
To further refine the system, we incorporate various result verification methods and adopt an incremental database construction approach. Through extensive ablation experiments, we rigorously assess the effectiveness of each component.
% Specifically, we leverage the effectiveness of dual-level retrieval and optimize its performance in a 32k context for maximum precision, and compare logic-based retrieval and dual-level retrieval to enhance overall functionality.
% Specifically, we integrate a dual-level retrieval method with a logic-form retrieval method.
Our implementation includes comparative experiments on SeedBench, where Qwen2.5-7B-Instruct initially underperformed. ROGRAG significantly improves the score from 60.0\% to 75.0\% and outperforms mainstream methods.
Experiments on domain-specific datasets reveal that dual-level retrieval enhances fuzzy matching, while logic form retrieval improves structured reasoning, highlighting the importance of multi-stage retrieval. 
% Furthermore, we propose a multi-stage verification mechanism to improve retrieval robustness without increasing computational cost.
% Empirical results show significant accuracy gains over baselines, highlighting the importance of adaptive retrieval. 
% ROGRAG is deployed on an online research platform\footnote{\url{https://seedllm.org.cn}} and released as an open-source resource\footnote{\url{https://github.com/tpoisonooo/ROGRAG}}. It supports easy installation with pip and a demo video\footnote{\url{https://www.youtube.com/watch?v=NL1_cuIiX-w}} is provided. 
ROGRAG is released as an open-source resource\footnote{\url{https://github.com/tpoisonooo/ROGRAG}} and supports installation with \texttt{pip}.
\end{abstract}

\begin{figure}[t]
    \centering 
    \includegraphics[width=\columnwidth]{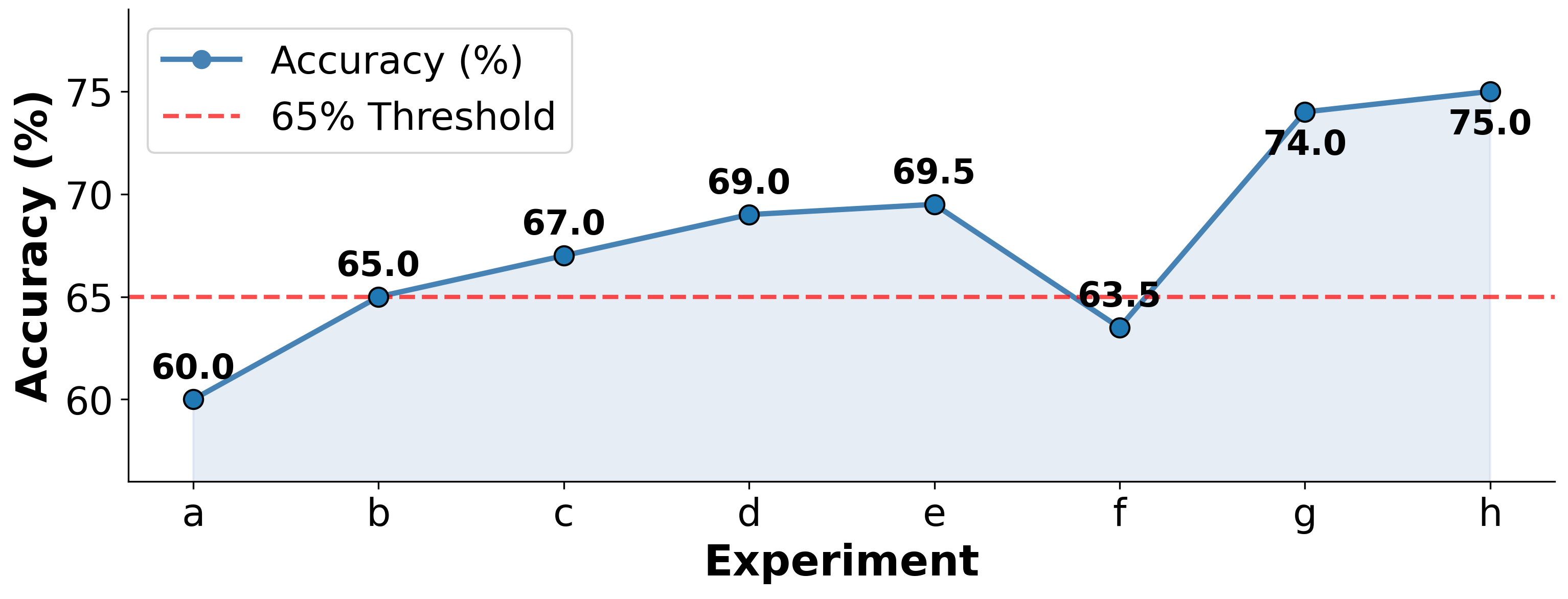}
    \caption{Performance improvements with each experiment – (a) Our initial system, (b) Remove abundant zero-shot example during retrieval, (c) Revert LLM $rope\_scaling$ default value, (d) Use 8k length for $nodes$ and $edges$, 12k length for $chunks$, (e) Expand low-level keys, (f) Exact matching method, (g) Revert to dual-level method and optimize NER prompt, (h) Fuse logic form retrieval with pre-check.}
    \label{fig:precision}
\end{figure}

\section{Introduction}

\begin{figure*}[t]
    \centering 
    \includegraphics[width=\textwidth]{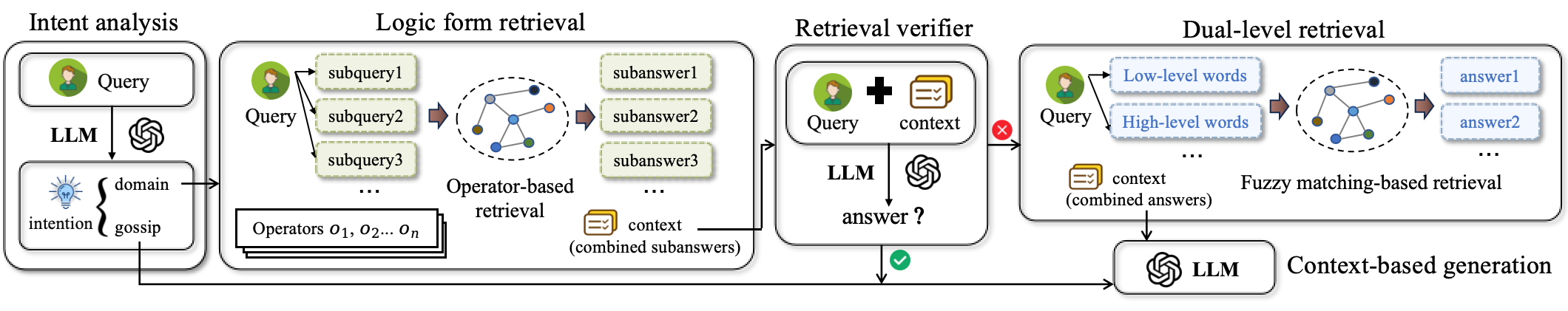}
    \caption{Multi-stage retrieval mechanism.
    User queries are first analyzed by an LLM to identify their intent and domain. The system then performs two retrieval strategies: logic form retrieval based on operator reasoning, and dual-level retrieval leveraging fuzzy matching. A verifier determines whether the retrieved context sufficiently answers the query. The final answer is generated by the LLM based on verified context.}
    
    % Receive user queries and process them using large language models, and then generate the final result after intent recognition, logical form generation, and dual-level retrieval.}
    \label{fig:pipeline}
\end{figure*}

% RAG is an artificial intelligence technique that integrates information retrieval with language generation models. It enhances the ability of large language models (LLMs) to handle knowledge-intensive tasks by retrieving relevant information from external knowledge bases and feeding it to the models as prompts. Typically, RAG consists of three stages: \textit{indexing}, \textit{retrieval}, and \textit{generation}. In its simplest form, indexing and retrieval rely solely on dense methods \cite{youdao_bcembedding_2023,  kong2024huixiangdou}, which match keyword features by extracting and comparing their distances. For example, the input sentences ``This board is great'' and ``This development board is terrible'' may have a similarity score as high as 0.7 due to their shared topic, despite their opposing semantics \cite{kexuefm-4122}. 
The rapid advancement of LLMs has significantly enhanced natural language processing (NLP) tasks \citep{min2023recent}. However, their reliance on finite training data and static pre-trained knowledge limits their effectiveness in knowledge-intensive applications, such as question answering (QA) and complex reasoning \cite{roberts2020much}. Retrieval-augmented generation (RAG) addresses these limitations by integrating information retrieval with language generation, improving factual accuracy and adaptability \cite{lewis2020retrieval}.

%To improve generation performance, GraphRAG methods introduce a combination of graph structures and dense methods during indexing and retrieval to represent the relationships between pieces of knowledge. For example, Microsoft's GraphRAG \cite{edge2024local} employs community detection algorithms during retrieval to answer uusermmarization questions. KAG \cite{liang2024kagboostingllmsprofessional} expands the knowledge graph into quads and decomposes query into sub-query to enhance logical coherence. LightRAG \cite{guo2024lightragsimplefastretrievalaugmented}, on the other hand, uses a dual-level approach to decompose the knowledge base and queries, thereby increasing the success rate of fuzzy matching.
Traditional RAG methods typically rely on dense retrieval \citep{karpukhin2020dense} or keyword-based matching to obtain relevant information in response to user queries. While effective for many tasks, these approaches often struggle with complex reasoning tasks that require understanding relationships between entities or synthesizing multi-hop knowledge. To overcome these limitations, recent research has explored GraphRAG, which incorporates structured knowledge representations such as knowledge graphs to enhance both retrieval accuracy and reasoning capability \cite{guo2024lightragsimplefastretrievalaugmented, liang2024kagboostingllmsprofessional}. By explicitly modeling entities and their relations, GraphRAG improves retrieval precision and facilitates structured reasoning.

%Among these three stages, indexing involves constructing a knowledge graph, often relying on Named Entity Recognition (NER) and requiring the repeated input of the entire knowledge base. This process leads to significant token consumption in LLMs. Therefore, analyzing the indexing process to clarify its cost-effectiveness is necessary. The retrieval stage depends on the accuracy of LLMs, such as decomposing queries into low-level and high-level representations to capture entities and relationships, or breaking down query for step-by-step execution. We need to verify the robustness of retrieval methods against LLM inaccuracies, ensuring that the retrieval results remain unaffected even if LLMs fail to correctly identify entities (e.g., ``Zhefu 80'').
%However, LLM training tokens have covered most publicly available RAG evaluation datasets. Due to the nature of LLMs, input prompts influence the probability distribution of next-token predictions \cite{vonoswald2023transformerslearnincontextgradient}, it is difficult to determine whether performance improvements stem from the GraphRAG method itself or from training data, as shown in \cref{tab:intro_llm_vs_gt}. 

Despite its potential, the development and evaluation of GraphRAG systems present several challenges. First, these pipelines typically consist of multiple interdependent components—including entity extraction, knowledge graph construction, query decomposition, retrieval mechanisms, and response generation \cite{lewis2020retrieval}—making it difficult to assess the contribution of each individual module. %Second, the widespread use of publicly available RAG benchmarks in LLM pretraining corpora complicates evaluation, 
Second, the widespread use of publicly available RAG benchmarks in LLM pretraining corpora complicates evaluation. As demonstrated in Table \ref{tab:intro_llm_vs_gt}, we verify that high performance may result from model memorization rather than true retrieval capability \citep{lewis2021paq}. Finally, many GraphRAG systems rely on heuristic-driven query decomposition, which may introduce errors if LLMs fail to generate accurate sub-queries, thereby degrading retrieval quality.

\begin{table}[t]
\centering
\setlength{\tabcolsep}{1.2pt}
\begin{tabular}{lccc}
\hline
\textbf{Domain} & \textbf{LLM win rate} & \textbf{Length ratio} \\\hline
Argriculture & 0.97 & \phantom{0}9.25 \\
Art & 0.91 & \phantom{0}8.20 \\
Biography & 0.82 & \phantom{0}7.69 \\
Cooking & 0.88 & \phantom{0}7.70 \\
Computer Science & 0.97 & 10.77 \\
Fiction & 0.65 & \phantom{0}6.63 \\
Finance & 0.85 & \phantom{0}9.33 \\
Health & 0.93 & \phantom{0}8.97 \\\hline

\end{tabular}
\caption{
% We execut the questions from the UltraDomain \cite{qian2024memoragmovingnextgenrag} dataset directly using Qwen2.5-7B-Instruct \cite{qwen2025qwen25technicalreport} then queried the Kimi API\footnote{See \url{https://platform.moonshot.cn}} for its preference between the LLM and the ground truth (GT). We calculated the LLM win rate based on preferences and also measured average token length ratio of LLM responses to GT. 
Validation of the use of the RAG benchmark in training models. The questions from the UltraDomain \cite{qian2024memorag} dataset are first executed using Qwen2.5-7B-Instruct \cite{qwen2025qwen25technicalreport}. The Kimi API\protect\footnotemark\ is then queried to compare the LLM's responses with the ground truth (GT) and determine its preference. Using these preferences, the LLM win rate is calculated, along with the average token length ratio between LLM's responses and the GT.
The results show that the direct responses from the 7B model significantly outperform GT, highlighting the difficulty of validating the RAG system's effectiveness on such datasets.}
\label{tab:intro_llm_vs_gt}
\end{table}
\footnotetext{See \url{https://platform.moonshot.cn} for Kimi API.}
% To clarify the effectiveness of each component of the GraphRAG pipeline, we selected three popular GraphRAG implementations (LightRAG, KAG, and DB-GPT \cite{xue2024dbgptempoweringdatabaseinteractions}, with a combined GitHub star count of 40k) and merged their codebases for ablation studies\footnote{Due to the immense workload (with over 18k lines of code), we could only evaluate the impact of methodological changes and could not ensure identical results before and after codebase integration. However, this does not affect the conclusions of this paper. Meanwhile, the high cost of manually selecting poorly performing LLM datasets limits our findings to a single domain}. To avoid issues related to LLMs being too powerful or having seen the RAG test sets, we manually selected domain-specific dataset where LLMs performed average. In summary, our contributions are as follows:

To systematically address these challenges, we introduce ROGRAG, an \textbf{R}obustly \textbf{O}ptimized \textbf{G}raph\textbf{RAG} framework designed for knowledge-intensive domains where the performance of \textit{vanilla} LLM remains suboptimal. Our method yields a substantial performance improvement, increasing the score from 60.0\% to 75.0\%, as illustrated in Figure \ref{fig:precision}. We improve the accuracy of the GraphRAG system by integrating and refining four seminal GraphRAG methodologies: DB-GPT \citep{xue2023db} (scalability), LightRAG~\citep{guo2024lightragsimplefastretrievalaugmented} (simple implementation), KAG \citep{liang2024kagboostingllmsprofessional} (reasoning), and HuixiangDou \citep{kong2024huixiangdou} (robustness), into a unified system. This integration not only leverages the advantages from different GraphRAG implementations, but also enables a comprehensive ablation study on the contributions of different indexing, retrieval, and generation strategies. 
% LightRAG - simple implementation
% KAG - logic-form retrieval
% DB-GPT-scaling up
% HXD - robustness
The system prioritizes query decomposition, and degrades to fuzzy matching if decomposition fails or verification is unsuccessful, as shown in Figure \ref{fig:pipeline}. This mechanism ensures robustness and enables continuous streaming responses. ROGRAG also retrains key components from the previous generation, including refusal-to-answer and intent slots. Additionally, we adopt domain-specific datasets where LLMs exhibit lower baseline scores to ensure that observed improvements reflect genuine retrieval enhancements rather than model memorization effects. Our main contributions are as follows:
\begin{itemize}[noitemsep, topsep=0pt, leftmargin=*]
    %\item We integrated four popular methods LightRAG, KAG, HuixiangDou and DB-GPT into a unified framework, merging over 18,000 lines of code. 
    \item \textbf{Unified GraphRAG Framework:} We merge multiple leading GraphRAG implementations into a single, extensible pipeline for structured retrieval and reasoning. Additionally, we introduce incremental database construction to dynamically expand and refine knowledge graphs.
    %\item We manually curated domain-specific datasets where LLMs underperform and conducted extensive parameter comparisons and ablation studies to evaluate their performance and interactions.
    \item \textbf{Enhanced Retrieval Mechanism:} We evaluate and refine retrieval techniques, including dual-level query decomposition, logic form retrieval, and fuzzy matching, with various result verification to improve accuracy and adaptability.
    %\item We open-sourced our implementation, \textit{ROGRAG}, to facilitate further research and development. 
    \item \textbf{Rigorous Empirical Evaluation:} Through comprehensive ablation studies on datasets where LLMs do not achieve trivial success, we provide insights into the key factors that contribute to performance improvements in GraphRAG-based QA systems. Additionally, the system is successfully deployed on the platform for user access.
\end{itemize}

% By open-sourcing ROGRAG, we aim to support further research and development in retrieval-augmented generation, particularly in applications requiring high-fidelity knowledge retrieval and reasoning. The remainder of this paper is structured as follows: Section 2 provides background on GraphRAG methodologies, Section 3 details our experimental setup, Section 4 analyzes indexing procedures, Section 5 evaluates retrieval and generation methods, and Section 6 presents the final optimized pipeline. Finally, Section 7 discusses conclusions and future research directions.

\section{Related Work}
\label{sec:related}

\subsection{Retrieval-Augmented Generation}
%Retrieval-augmented generation (RAG) has attracted widespread attention as a technique to enhance LLMs by integrating external knowledge retrieval \cite{gao2023retrieval,fan2024survey}. RAG \cite{lewis2020retrieval} includes three parts: indexing, retrieval\citep{robertson2009probabilistic,karpukhin2020dense} and generation. It searches for other knowledge, skills and tools based on semantic similarity \citep{khattab2020colbert} of the user-entered query. 
%In recent years, there have been some studies on RAG. \citep{borgeaud2022improving} proposed RETRO, a language model that integrates a large-scale retrieval mechanism during training and reasoning to improve the generation ability of language models and reduce hallucinations. \citep{cheng2024lift} introduced a self-memory mechanism, allowing the model to use its own output as a retrieval source during the generation process, thereby improving the quality of text generation.
%However, the traditional RAG model has certain limitations. RAG is weak in processing structured data because it relies on single document retrieval and cannot grasp global information, while data often have complex, multi-hop connections.
RAG enhances LLMs by integrating external knowledge retrieval to improve factual accuracy and contextual relevance \cite{gao2023retrieval, fan2024survey}. The standard RAG pipeline \cite{lewis2020retrieval} consists of three key components: indexing, retrieval, and generation. Retrieval methods typically rely on semantic similarity \cite{khattab2020colbert} to identify relevant knowledge from external sources. Recent advancements in RAG have focused on mitigating hallucinations and improving generation quality. For instance, RETRO \cite{borgeaud2022improving} employs large-scale retrieval during both training and inference, while Lift-RAG \cite{cheng2024lift} introduces self-memory mechanisms to leverage generated content for subsequent retrieval. However, traditional RAG models struggle with structured data, as they primarily rely on single-document retrieval and fail to capture complex multi-hop relationships.

\subsection{Graph Retrieval-Augmented Generation}
%GraphRAG is a framework that combines graph neural networks (GNNs)\citep{zhou2020graph} and retrieval-augmented generation (RAG). It provides a more structured graph structure (such as knowledge graphs Freebase\cite{bollacker2008freebase} and Wikidata\cite{vrandevcic2014wikidata}) to model the relationship between entities. Unlike traditional RAG models, GraphRAG retrieves task-related entity and relationship information from graph databases, rather than relying solely on the content in the text corpus. For example, K-BERT \cite{liu2020k} uses entity and relationship information in knowledge graphs to enhance the representation of word vectors. Traditional graph data representation learning methods usually rely on complex graph convolution operations such as graph convolutional networks (GCNs)\cite{kipf2016semi} to capture the relationship between nodes. Graph-BERT \cite{zhang2020graph} proposes that learning graph structured data can use self-attention mechanisms\citep{vaswani2017attention} to capture the relationship between nodes like learning text data, replacing traditional graph convolution operations.
To address the limitations of conventional RAG, GraphRAG integrates graph-based structures, including knowledge graphs like Freebase \cite{bollacker2008freebase} and Wikidata \cite{vrandevcic2014wikidata}, into the retrieval process. By leveraging entity-relationship graphs, GraphRAG provides richer contextual information to enhance both retrieval and generation tasks. 
% Approaches such as K-BERT \cite{liu2020k} incorporate knowledge graphs into transformer models to refine word representations. Traditional graph-based methods, including graph convolutional networks \cite{kipf2016semi}, capture relational dependencies through iterative message passing. More recent architectures, such as Graph-BERT 
% \cite{zhang2020graph}, replace graph convolutions with self-attention mechanisms \cite{vaswani2017attention}, enabling more flexible and scalable representation learning.
Since its introduction \cite{edge2024local}, researchers have explored how integrating graph-structured data improves the model’s ability to capture complex dependencies \cite{han2024retrieval}. Meanwhile, a systematic analysis of the application of GraphRAG in customizing LLMs has also been conducted \cite{zhang2025survey}.

Despite these significant advancements, existing GraphRAG models still face challenges in balancing retrieval accuracy, computational efficiency, and adaptability to diverse query structures \citep{peng2024graph}. Our work builds upon these foundations by refining GraphRAG techniques to improve retrieval precision and response coherence while maintaining efficient knowledge integration.

\section{Methodology}
% This section provides an overview of the GraphRAG approach, detailing its architectural components and key methodological choices. We describe the indexing, retrieval, and generation processes, followed by a discussion of the datasets used for evaluation.
% This section details the GraphRAG architectural components and key methodological choices. An algorithmic overview of the GraphRAG approach is detailed in Algorithm \ref{algo:graphrag} in
% Appendix A.1.
% The GraphRAG framework is structured around three key subtasks: \textbf{indexing}, \textbf{retrieval}, and \textbf{generation}. These components jointly determine the system's overall performance.

In this section, we describe the components of GraphRAG, including indexing, retrieval, and generation, as well as the key methodological choices. A comprehensive algorithmic overview can be found in Algorithm \ref{algo:graphrag} in Appendix \ref{sec:A.1}.

\begin{figure}[t]
    \centering
    \includegraphics[width=\columnwidth]{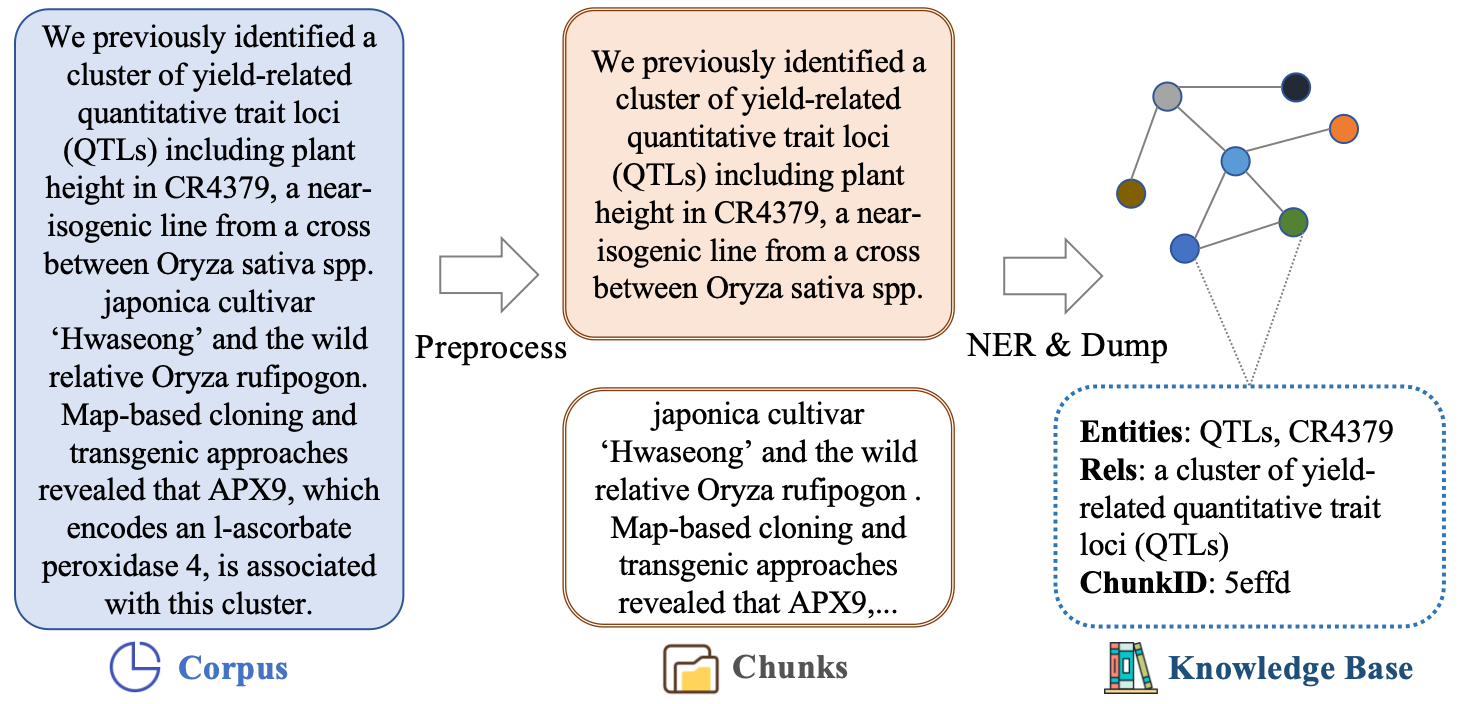}
    \caption{Architecture of GraphRAG indexing. The raw corpus is first cleaned and segmented into manageable chunks. Entities, relationships, keywords and descriptions are then extracted from each chunk. Subsequently, graph nodes and edges are constructed and linked back to their corresponding text chunks.}
    % , enabling a structured representation of the underlying knowledge.}
    \label{fig:indexing}
\end{figure}

\subsection{Graph-Based Indexing}
% The indexing process, denoted by \(f\) in Algorithm \ref{algo:graphrag}, consists of the following stages:
The indexing process, represented by \(f\) in Algorithm \ref{algo:graphrag}, consists of the preprocess, named entity recognition (NER), and dump stages. An overview of this indexing workflow is shown in Figure \ref{fig:indexing}.

% \begin{itemize}[noitemsep, topsep=0pt, leftmargin=*]
\noindent\textbf{Preprocess.} %Standardizes corpus files and segments them into discrete text chunks.
Corpus files from heterogeneous sources are standardized and segmented into discrete text chunks to ensure structural uniformity.

\noindent\textbf{NER.} 
%Identifies and extracts structured triplets in the form of \(\langle \text{entity}, \text{relation}, \text{description} \rangle\), then constructs nodes and edges of a graph, capturing multi-hop dependencies within the corpus.
% Structured \(\langle \text{entity}, \text{relation}, \text{description} \rangle\) triplets are identified and extracted from segmented text. Subsequently, a graph is constructed by establishing node-edge relationships to capture multi-hop corpus dependencies across the corpus.
\(\langle \text{entity}_s, \text{relation}, \text{entity}_o \rangle\) triplets are initially extracted from segmented text, along with corresponding keywords, description and weight (\textit{e.g.}, $\langle \text{Marie Curie}, \text{discovered}, \text{radium} \rangle$; scientist, discovery; Marie Curie discovered radium in 1898; 4.5). Subsequently, a graph is constructed by establishing node-edge relationships to capture complex multi-hop dependencies across the corpus.

\noindent\textbf{Dump.} The extracted entities, relations, and their corresponding embeddings are stored in a structured database for efficient retrieval and analysis.
% \end{itemize}

% Formally, this process can be expressed as:

% \begin{equation}
%     f = \text{store\_data}\bigl(\text{extract\_entities}\bigl(\text{preprocess}(C)\bigr)\bigr).
% \end{equation}
% An illustration of this indexing workflow is provided in Figure \ref{fig:indexing}.

% \begin{figure}[t]
%     \centering
%     \includegraphics[width=\columnwidth]{ROGRAG_indexing.png}
%     \caption{Architecture of GraphRAG indexing. The raw corpus is first cleaned and split into chunks, followed by the extraction of entities, keywords, relationships, and descriptions, and finally, the graph nodes and edges are linked to the chunks.}
%     \label{fig:indexing}
% \end{figure}

\subsection{Graph-Guided Retrieval}
The retrieval phase is governed by both \(g\) and the iterative selection in Algorithm~\ref{algo:graphrag}. There are two main methods: dual-level and logic form.

\begin{figure}[t]
    \centering 
    \includegraphics[width=\columnwidth]{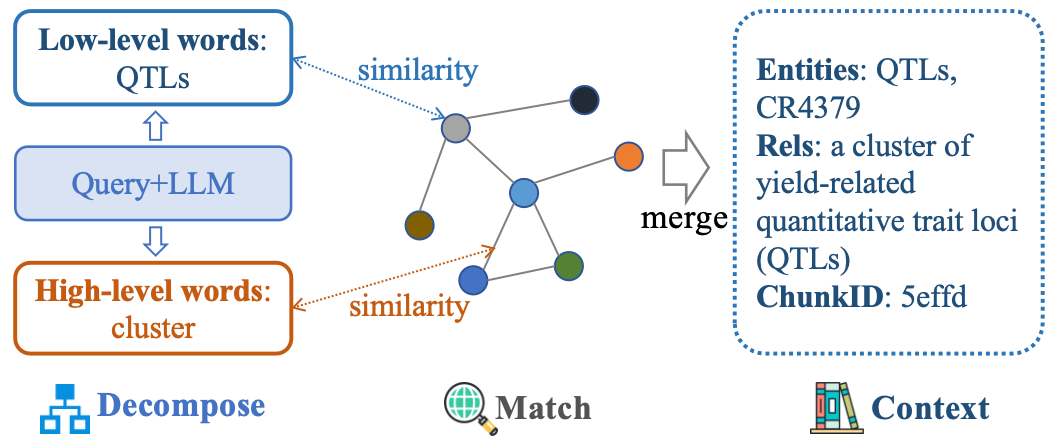}
    \caption{Dual-level retrieval method. User query would be decomposed into low-level and high-level keywords, then match with the knowledge graph.}
    \label{fig:dual-level retrieval}
\end{figure}

% \begin{itemize}[noitemsep, topsep=0pt, leftmargin=*]
    \noindent\textbf{Dual-Level Method.} As illustrated in Figure~\ref{fig:dual-level retrieval}, the query is decomposed into two components: (i) low-level keywords representing entities and (ii) high-level relational descriptions. Entities are identified and matched to corresponding nodes in the graph, often using fuzzy matching, and their associated edges are subsequently retrieved. Similarly, relational keywords are mapped to edges to retrieve connected nodes. The retrieved results are then merged, with redundant edges, nodes, and chunk references systematically removed to refine the final retrieval context. This approach leverages multi-granularity features for layered fuzzy matching, improving retrieval coverage on ill-formed or complex queries and enhancing robustness.
    
    \noindent\textbf{Logic Form Method.} Inspired by knowledge-aware reasoning frameworks, this approach utilizes a predefined set of operators (\textit{e.g.}, filtering, aggregation) to decompose complex queries. An LLM is employed to transform natural language queries into a structured sequence of retrieval operations, which are iteratively refined to enhance the retrieval context. The pseudocode for this approach is presented in Algorithm \ref{algo:logic form retrieval} in Appendix \ref{sec:A.2}.
% \end{itemize}

% \begin{algorithm}[ht]
% \caption{Logic form retrieval}
% \begin{algorithmic}[1]
% \State \textbf{Input:} Operator set \( O = \{o_1, o_2, \dots, o_n\} \), where each \( o_i = (\text{operator}_i, \text{function}_i) \)
% \State \textbf{Input:} User query \( Q \)
% \State \textbf{Output:} History

% \State Decompose query \( Q \) into a list of subqueries \( L \) using LLM
% \State \( L \gets \{(q_1, a_1), (q_2, a_2), \dots, (q_m, a_m)\} \), where each \( a_j \in O \)

% \For{\((q_j, a_j) \in L\)}
%     \State Identify the corresponding operator \( o_j \) for \( a_j \)
%     \State Execute \( a_j \gets o_j(q_j) \)
% \EndFor

% \State Concatenate all sub-queries and sub-answers \( a_j \) into history
% \State \textbf{return} History
% \end{algorithmic}
% \label{algo:logic form retrieval}
% \end{algorithm}

% \begin{figure}[t]
%     \centering 
%     \includegraphics[width=\columnwidth]{ROGRAG_dual_retrieval.png}
%     \caption{Dual-level retrieval method. User query would be decomposed into low-level and high-level keywords, then match with knowledge graph.}
%     \label{fig:dual-level retrieval}
% \end{figure}

\subsection{Graph-Enhanced Generation}
%Most of RAG utilize LLMs for both generation and evaluation. For example, input formatting and prompt engineering are used to optimize the generation process, while the LLM assesses whether the generated response answers the question and whether the context is relevant to the query.
During the generation stage, most RAG architectures leverage LLMs to (i) format and present the retrieved context as part of a prompt, (ii) produce the final response, and (iii) evaluate or verify whether the generated output correctly addresses the query. Techniques such as input formatting, prompt engineering, and self-consistency checks are often employed to optimize these generations.

\section{Experiment}
In this section, we describe the experimental setup and overall result of our GraphRAG system.

\subsection{Experimental Setup}
\noindent\textbf{Methods and Models.} We adopt techniques from HuixiangDou~\cite{kong2024huixiangdou} regarding the refusal-to-answer task. To switch between knowledge bases, we use TuGraph \cite{TuGraph} for knowledge graph storage and BCEmbedding \cite{youdao_bcembedding_2023} to extract features for entities and relations due to its effectiveness in generating high-quality embeddings. 
% With a maximum token size of 512, the extracted features are subsequently indexed in Faiss \cite{douze2024faisslibrary}. 
The extracted features are subsequently indexed in Faiss \cite{douze2024faisslibrary}.
To ensure that our experiments require minimal computational resources, we conduct tests on Qwen2.5-7B-Instruct, a relatively lightweight model that offers a good balance of efficiency and capability. Appendix \ref{sec:B} provides further details.

\renewcommand{\arraystretch}{1.2}
\begin{table}[t]
\centering
\small
\setlength{\tabcolsep}{3pt}
\begin{tabular}{lccccc}
\hline
% \textbf{Subset} & \textbf{Metric} & \textbf{LLM baseline} & \textbf{ROGRAG} \\\hline
% QA-1 & Accuracy & 0.57 & \textbf{0.75} \\
% QA-2 & F1 Score & 0.71 & \textbf{0.79} \\
% QA-3 & Rouge & 0.16 & \textbf{0.36} \\
% QA-4 & Rouge & 0.35 & \textbf{0.38} \\\hline
\multirow{3}{*}{\textbf{Method}} & \multicolumn{4}{c}{\textbf{SeedBench Subsets}} \\ \cline{2-5}
& \textbf{QA-1} & \textbf{QA-2} & \textbf{QA-3} & \textbf{QA-4} \\ 
& (Accuracy) & (F1) & (Rouge) & (Rouge) \\\hline
\textit{vanilla} (\textbf{w/o RAG}) & 0.57 & 0.71 & 0.16 & 0.35 \\
\textbf{LangChain} & 0.68 & 0.68 & 0.15 & 0.04 \\
\textbf{BM25} & 0.65 & 0.69 & 0.23 & 0.03 \\
\textbf{RQ-RAG} & 0.59 & 0.62 & 0.17 & 0.33 \\\hline
\textbf{ROGRAG (Ours)} & \textbf{0.75} & \textbf{0.79} & \textbf{0.36} & \textbf{0.38} \\\hline
\end{tabular}
\caption{A comparison of the test scores between several mainstream RAG systems and our proposed system. The experiment is conducted using Qwen2.5-7B-Instruct on SeedBench, while the BM25~\cite{FlashRAG} and RQ-RAG~\cite{chan2024rq} methods are implemented based on FlashRAG \cite{FlashRAG}. The GraphRAG framework, ROGRAG, which we proposed, outperforms all other methods across all four subsets of SeedBench, demonstrating the effectiveness of the subsequent optimizations.}
% After ablation studies on the baseline, we present some results tested with the new version on SeedBench.}
\label{tab:ROGRAG}
\end{table}

\noindent\textbf{Data and Evaluation.} We adopt the SeedBench \cite{ying2025seedbench} dataset, a domain-specific benchmark curated by human experts to evaluate systems. 
% For a detailed introduction to other public datasets we use, please see Appendix \ref{sec:B}.
In subsequent ablation studies, we applied multiple-choice questions from QA-1 subset of SeedBench, with accuracy as evaluation metric.

\subsection{Overall Result}
Table~\ref{tab:ROGRAG} compares the test scores of several mainstream RAG systems and our proposed system. Initially, we evaluate the large model's direct response without RAG (\textit{vanilla}) as baselines and then combine the four foundational GraphRAG methodologies to construct our initial system. Next, we conduct detailed ablation experiments on different components of the system, comparing various methods and parameters to improve each component, thereby progressively enhancing the test outcomes. Finally, we introduce ROGRAG and compare it with three mainstream RAG systems, including inverted indexing, similarity-based retrieval, and multi-round answering techniques. 

Our experiments show that ROGRAG outperforms all other systems and the baseline across all four subsets of SeedBench. Additionally, it also proves that Qwen2.5-7B-Instruct is initially unsuitable for these tasks and subsequent optimizations are effective. Interestingly, on the QA-2 dataset, the non-RAG approach even outperforms mainstream methods. This suggests that RAG responses could be misled by irrelevant context and applying RAG does not always lead to higher accuracy, particularly in domains where LLMs lack familiarity. The poor performance of LangChain \citep{Chase_LangChain_2022} and BM25 \citep{FlashRAG} methods on the QA-4 generation task may be due to parameter settings that resulted in minimal relevant content being retrieved, limiting the model's ability to generate accurate answers.

% \begin{table}[t]
% \centering
% \begin{tabular}{lcc}
% \hline
% \textbf{Subset} & \textbf{Metric} & \textbf{Score} \\\hline
% % \textit{QA-1} & Accuracy & 0.57 \\
% % \textit{QA-2} & F1 Score & 0.71 \\
% % \textit{QA-3} & Rouge & 0.16 \\
% % \textit{QA-4} & Rouge & 0.35 \\
% QA-1 & Accuracy & 0.57 \\
% QA-2 & F1 Score & 0.71 \\
% QA-3 & Rouge & 0.16 \\
% QA-4 & Rouge & 0.35 \\
% \hline

% \end{tabular}
% \caption{Baseline results of Qwen2.5-7B-Instruct on our domain-specific data. The poor performance of LLM on this data can demonstrate that the subsequent optimization of GraphRAG is effective. Score is the value corresponding to the metric.
% % F1 is the harmonic mean of precision and recall, while rouge measures the overlap of n-grams between generated and reference texts.
% }
% \label{tab:qwen_baseline}
% \end{table}

\section{Indexing Analysis}

In this section, we begin by introducing different NER strategies and then proceed to validate the parameters associated with indexing.

% \subsection{How Different Loop NER Methods Affect the Precision?}
\begin{table}[t]
\centering
\setlength{\tabcolsep}{4pt}
\begin{tabular}{lccc}
\hline
\textbf{Version} & \textbf{Nodes} & \textbf{Edges} & \textbf{Accuracy} \\\hline
Trial & 20,739 & 19,857 & 0.61 \\
Base & 21,838 & 26,847 & 0.69 \\
Optimize Prompt & 29,086 & 35,750 & 0.74 \\\hline 
\end{tabular}
\caption{
% Relationship between the number of nodes and edges in a graph and the accuracy under different loop NER implementations. 
Number of nodes and edges generated by different Loop NER Strategies and the methods’ impact on accuracy. Increasing the quantity can improve accuracy.}
% , as erroneous entity words generated by LLMs become isolated nodes and do not affect the overall graph structure and final accuracy.}
\label{tab:ner_bench}
\end{table} 

\subsection{Loop NER Strategies}
To maximize the recall of the NER, GraphRAG tends to iteratively call the LLM in order to extract as many entities as possible. The common stopping condition is based on the LLM's judgment of whether any entities have been missed. We compare two implementations of iterative NER, as shown in Algorithm \ref{algo:ner} in Appendix \ref{sec:A.3}. The trial version (trial) follows a standard procedural logic: it performs NER first, then queries the LLM whether further extraction is needed, and proceeds only if the response is affirmative. In contrast, Baseline version (base) deviates from this flow. After performing NER, it informs the LLM that more entities may exist and requests further extraction before entering the if-condition.

As shown in Table \ref{tab:ner_bench}, the larger nodes and edges in the knowledge graph is positively correlated with higher precision, and the erroneous entities generated by LLM have minimal impact on the overall graph structure and final accuracy.  Therefore, we can optimize the prompt by specifying entity types and splitting examples to improve accuracy. 

% \begin{itemize}[noitemsep, topsep=0pt, leftmargin=*]
%     \item More entity results improve precision, thus we can optimize the prompt by specifying entity types and splitting examples
%     % \footnote{The optimized prompt is provided in \url{https://github.com/tpoisonooo/HuixiangDou2/blob/main/huixiangdou/service/prompt/kag.py}.}.
%     \item Smaller \textit{chunk size} may yield better results.
% \end{itemize}

\subsection{Max LLM Context Length}
While LLMs typically perform well in needle-in-a-haystack experiments, RAG systems often need to focus more on subtle and implicit expressions within the corpus. To extend the maximum input length in YaRN \cite{peng2023yarn}, we modify the parameter $rope\_scaling$ , which modifies the rotary position embedding (RoPE) scaling strategy in order to accommodate longer contexts. Our empirical results show that increasing the length from 32k to 64k leads to a noticeable 2\% drop in precision. Therefore, a smaller $rope\_scaling$ setting is generally preferable, as shown in Table \ref{tab:llm_bench}. Similarly, reducing \textit{chunk size} may lead to better results, as it allows for more accurate information extraction.

\begin{table}[t]
\centering
\begin{tabular}{ccc}
\hline
\textbf{Max Length} & \textbf{Accuracy} \\\hline
32k & 0.67 \\
64k & 0.65 \\\hline 
\end{tabular}
\caption{Impact of different maximum context lengths of LLM on accuracy. When the model's maximum length is sufficient, a smaller $rope\_scaling$ is preferred.}
\label{tab:llm_bench}
\end{table}

\section{Retrieval \& Generation Analysis} 

In this section, we first validate the parameters that impact performance, followed by a comparison of different retrieval and verification methods. 
% and ultimately integrate the two approaches to achieve the optimal effect.

\subsection{Representation and Matching Granularity}

Based on the previous conclusions, we hypothesize that the essence of the dual-level method lies in approximate matching. The richer the representations derived from the corpus and query, the higher the overall precision. To validate this, we conducted experiments in two opposite directions.

\noindent\textbf{Extended Queries and Low-Level Keys.} Since LLMs often struggle with domain-specific data (\textit{e.g.}, mistaking entities for relationships), we expand the maximum length of the query representation ($R_q$ in Algorithm \ref{algo:graphrag}) and increase the number of low-level keys. This approach aims to allow the query to incorporate more detailed information, thereby enhancing the accuracy of matching.

\noindent\textbf{Exact Matching.} Instead of concatenating the entity list, we independently store the features of each entity during the indexing phase and match individual entities during query time. This method aims to improve precision by avoiding the noise typically introduced by approximate matching.

Our hypotheses are validated in Table \ref{tab:repr_length}. Increasing the maximum length of $R_q$ to 28k results in a 4\% improvement in precision compared to baseline. Expanding the number of low-level keys helps fix some rare bad cases. In contrast, switching to exact matching leads to a decrease in precision. This suggests that fuzzy matching is essential for capturing the nuances of the query, as exact matching fails to account for the implicit keywords derived from the query. This aligns with common sense, as exact matching is too rigid for complex queries. 

\begin{table}[t]
\centering
\begin{tabular}{lc}
\hline
\textbf{Method} & \textbf{Accuracy} \\\hline
Dual-level & 0.650 \\
\;\text{+28k context length} & 0.690 \\
\;\;\text{+expand low-level keys} & 0.695 \\
Exact matching method & 0.635 \\\hline 

\end{tabular}
\caption{Validations of dual-level retrieval method in two opposite directions. 
% We hypothesize that the length of $R_q$ and $R_c$ improve the accuracy. Conducting validations in two opposite directions, the results supported this hypothesis.}
It is hypothesized that increasing the length of $R_q$ and $R_c$ will lead to improved accuracy, and the results support this hypothesis.}
\label{tab:repr_length}
\end{table}

\subsection{Dual-Level vs. Logic Form}
We compare the results of dual-level and logic form methods in Table \ref{tab:retrieval_method}. Although the dual-level method achieves higher precision, it fails to provide convincing answers to questions that require calculations (\textit{e.g.}, ``How much taller is Zhefu 802 than its parent?''). In our real-world scenario evaluations, responses generated by the logic form method are more concise and exhibit a clearer logical progression, preferred by domain experts.

\begin{table}[t]
\centering
\setlength{\tabcolsep}{5pt}
\begin{tabular}{lcc}
\hline
\textbf{Method} &  \textbf{Avg length} & \textbf{Accuracy} \\\hline
Optimized dual-level & 9863 & 0.74 \\
Logic form & 1699 & 0.55 \\\hline 

\end{tabular}
\caption{Average output context length of dual-level and logic form and methods' impact on accuracy. 
% The average lengths of the contexts generated by the two methods are tallied, and these retrieval results are used to answer SeedBench questions. 
Although the logic form retrieval method shows suboptimal precision, it provides higher information density.}
\label{tab:retrieval_method}
\end{table}

% If we had a verifier with 100\% accuracy, we could combine the strengths of both methods to achieve an accuracy of 0.77.

\subsection{Argument Checking vs. Result Checking}
RAG systems often employ LLM to verify the correctness of results. We compare two approaches: argument checking and result checking. The argument checking verifies whether the provided context can answer the question before generating the final response, while the result checking examines the question, context, and response together for overall coherence. The pseudocode is shown in Algorithm \ref{algo:check} in Appendix \ref{sec:A.4}. The results are summarized in Table \ref{tab:check}, which shows that argument checking is preferred. From the aspect of inference, the response diverts part of the LLM's attention, thereby reducing its ability to focus on the core question. From the aspect of model, Qwen-2.5-7B-Instruct is a causal model, where correct reasoning within the context typically leads to correct results. Therefore, result checking is redundant.

\begin{table}[t]
\centering
\begin{tabular}{lc}
\hline
\textbf{Method} & \textbf{Accuracy} \\\hline
% post-check & 0.745 \\
% pre-check & 0.715 \\\hline 
% Post-check & 0.75 \\
% Pre-check & 0.72 \\\hline 
Argument Checking & 0.75 \\
Result Checking & 0.72 \\\hline 
\end{tabular}
\caption{Performance comparison of checking strategies on logic form retrieval method. The results indicate that the argument checking yields better performance.}
\label{tab:check}
\end{table}

% \section{ROGRAG}
% Based on the ablation and analysis, we ultimately constructed the ROGRAG retrieval pipeline, as shown in Figure \ref{fig:pipeline}.It is available on GitHub at \footnote{\url{https://github.com/tpoisonooo/ROGRAG}} , and easily installable using pip according to this instruction \footnote{\url{https://github.com/tpoisonooo/ROGRAG/blob/main/docs/zh_cn/doc_how_to_run.md}}.

% \begin{figure}[t]
%     \centering 
%     \includegraphics[width=\columnwidth]{ROGRAG_pipeline.png}
%     \caption{Multi-stage retrieval.
%     User queries are received and processed using large language models, and the final result is generated after intent recognition, logical form generation, and dual-level retrieval.}
    
%     % Receive user queries and process them using large language models, and then generate the final result after intent recognition, logical form generation, and dual-level retrieval.}
%     \label{fig:pipeline}
% \end{figure}

% Its precision results are shown in Table \ref{tab:ROGRAG} 
% and features include: 
% \paragraph{Multi-Stage} Prioritizing query decomposition, and degrading to fuzzy matching if decomposition fails or verification does not pass. This mechanism ensures that the pipeline always supports streaming responses.
% \paragraph{Version Compatibility} ROGRAG retrains the useful refusual-to-answer and intent slots from the previous generation.

\section{Conclusion}
% Based on our engineering practice, we observe that the Logic Form, with its step-by-step approach, is more readily accepted by human domain experts. This method aligns well with human reasoning processes and provides clear, logical answers that are easier to interpret and validate.

% However, several challenges remain. First, constructing effective supervision signals, such as a high-accuracy Verifier, is difficult but crucial for further improving precision. Second, the initial steps decomposed by LLMs often deviate significantly from the knowledge graph, leading to failures in execution and result retrieval. These issues highlight the gap between LLM-generated plans and practical knowledge graph operations.

% Future work will focus on addressing these challenges. We plan to explore more robust methods for building verifiers and improving the alignment between LLM-generated steps and knowledge graph structures. Additionally, we aim to integrate hybrid approaches that combine the strengths of both Logic Form and Dual Level methods to achieve better overall performance. For a more detailed discussion on Experimental Section, please refer to Appendix \ref{sec:C}.
In this work, we introduce ROGRAG, a robustly optimized GraphRAG framework that addresses the limitations of existing retrieval-augmented generation pipelines in handling specialized and domain-specific queries. By combining dual-level and logic form methods in a multistage retrieval process, ROGRAG enhances retrieval robustness. The system is further enhanced with result verification methods and an incremental knowledge graph construction strategy. The empirical studies show that the logic form method, with its step-by-step approach, is more acceptable by domain experts. This method aligns well with human reasoning and provides clear and logical answers that are easy to interpret and validate. Lastly, future work shall focus on building a high-accuracy verifier, refining LLM decomposition steps, and exploring further enhancements to improve overall performance. For a more detailed discussion, see Appendix \ref{sec:C}.

\section*{Limitations}
Due to the large workload, we have only used Qwen2.5-7B-Instruct to conduct experiments on a single domain-specific dataset for the time being. In the future, we will explore the application of ROGRAG to more general models, test its performance on a broader range of domain-specific datasets, and enhance its robustness. However, developing the GraphRAG system presents several challenges, such as the difficulty of constructing an effective high-accuracy verifier, which is essential for further improving precision. On the other hand, due to the inherent limitations of large models and the noisy or incomplete corpus provided, errors in entity extraction, query decomposition, or knowledge retrieval within the GraphRAG method may propagate throughout the system, exacerbating inaccuracies in response generation. In addition, heuristic-driven query decomposition remains a potential bottleneck, as errors in subquery formation may degrade retrieval performance.

\section*{Acknowledgments}
This work was supported by Shanghai Artificial Intelligence Laboratory. The authors would like to thank Chenyang Wang from Huawei Technologies Co., Ltd and Zhe Ma from Shanghai Artificial Intelligence Laboratory for academic support.
% \clearpage
% \clearpage
\bibliography{anthology,custom}
\bibliographystyle{acl_natbib}

% \appendix

% \section{Logic Form Retrieval}

% \begin{algorithm}
% \caption{Logic form retrieval}
% \begin{algorithmic}[1]
% \State \textbf{Input:} Operator set \( O = \{o_1, o_2, \dots, o_n\} \), where each \( o_i = (\text{operator}_i, \text{function}_i) \)
% \State \textbf{Input:} User query \( Q \)
% \State \textbf{Output:} History

% \State Decompose query \( Q \) into a list of subqueries \( L \) using LLM
% \State \( L \gets \{(q_1, a_1), (q_2, a_2), \dots, (q_m, a_m)\} \), where each \( a_j \in O \)

% \For{\((q_j, a_j) \in L\)}
%     \State Identify the corresponding operator \( o_j \) for \( a_j \)
%     \State Execute \( a_j \gets o_j(q_j) \)
% \EndFor

% \State Concatenate all sub-queries and sub-answers \( a_j \) into history
% \State \textbf{return} History
% \end{algorithmic}
% \label{algo:logic form retrieval}
% \end{algorithm}

\clearpage
\appendix

\section{Additional Details on Methodology}
\label{sec:A}
\subsection{GraphRAG}
\label{sec:A.1}
 
Let the corpus be denoted by $C$ and the user query by $Q$. The GraphRAG framework, defined as $\mathsf{graphrag} = (f, g, d)$, consists of three primary functions:

\begin{itemize}[noitemsep, topsep=0pt, leftmargin=*]
    \item \textbf{Indexing function} $f$, which extracts structured representations from $C$, yielding $R_c$.
    \item \textbf{Retrieval function} $g$, which derives representations from $Q$, producing $R_q$.
    \item \textbf{Graph-based augmentation function} $d$, which iteratively constructs paths linking $R_c$ and $R_q$, refining the retrieval context.
\end{itemize}

\begin{algorithm}[ht]
\caption{GraphRAG}
\begin{algorithmic}[1]
\State Compute corpus representations: $R_c = f(C)$
\State Compute query representations: $R_q = g(Q)$
\State Initialize the retrieval set:  $S_0 \subseteq R_c$
\For{$k = 1, 2, \ldots$}
    \State Select the most relevant element: $e_k^+ = \arg\min_{e \in R_c \setminus S_{k-1}} \text{dist}(S_{k-1} \cup \{ e \}, R_q)$
    \State Remove the least relevant element: $e_k^- = \arg\min_{e \in S_{k-1}} \text{dist}(S_{k-1} \setminus \{ e \}, R_q)$
    \State Update retrieval set: $S_k = S_{k-1} \cup \{ e_k^+ \} \setminus \{ e_k^- \}$
    \If{$\text{dist}(S_k, R_q)$ does not improve}
        \State Terminate retrieval process
    \EndIf
\EndFor
\end{algorithmic}
\label{algo:graphrag}
\end{algorithm}

The indexing and retrieval processes are formalized in Algorithm \ref{algo:graphrag}. At each iteration, an element \(e_k^+\) is selected for inclusion if it minimizes the retrieval distance \(\text{dist}(S_{k-1}, R_q)\), while an element \(e_k^-\) is removed if it similarly optimizes the retrieved set. The stopping criterion ensures termination when no further improvements can be achieved.

\subsection{Logic Form Retrieval}
\label{sec:A.2}
Logic Form method. Inspired by knowledge-
aware reasoning frameworks, this approach employs a predefined set of operators (e.g., filtering, aggregation) to decompose complex queries.

\begin{algorithm}[ht]
\caption{Logic Form Retrieval}
\begin{algorithmic}[1]
\State \textbf{Input:} Operator set \( O = \{o_1, o_2, \dots, o_n\} \), where each \( o_i = (\text{operator}_i, \text{function}_i) \)
\State \textbf{Input:} User query \( Q \)
\State \textbf{Output:} History

\State Decompose query \( Q \) into a list of subqueries \( L \) using LLM
\State \( L \gets \{(q_1, a_1), (q_2, a_2), \dots, (q_m, a_m)\} \), where each \( a_j \in O \)

\For{\((q_j, a_j) \in L\)}
    \State Identify the corresponding operator \( o_j \) for \( a_j \)
    \State Execute \( a_j \gets o_j(q_j) \)
\EndFor

\State Concatenate all sub-queries and sub-answers \( a_j \) into history
\State \textbf{return} History
\end{algorithmic}
\label{algo:logic form retrieval}
\end{algorithm}

\subsection{Loop NER}
\label{sec:A.3}
Two implementations of iterative NER, including a trial version and a base version.

\begin{algorithm}[H]
\caption{Loop NER}
\begin{algorithmic}[1]
\State Initialize input text $T$
\State Initialize maximum attempts $MAX$
\State Initialize history $H \gets NER\_init(T)$
\State
\Function{trial}{}
    \For{$i = 0$ \textbf{to} $MAX$}
        \State $continue \gets LLM\_judge(H)$ 
        \If{$continue == \text{"no"}$}
            \State \textbf{break}
        \EndIf
    
        \State $result \gets NER\_continue(T, H)$
        \State $H \gets H \cup result$
    \EndFor
\EndFunction
\State
\Function{base}{} 
    \For{$i = 0$ \textbf{to} $MAX$}
        \State $result \gets NER\_continue(T, H)$
        \State $H \gets H \cup result$
        
        \State $continue \gets LLM\_judge(H)$
        \If{$continue == \text{"no"}$}
            \State \textbf{break}
        \EndIf
    \EndFor
\EndFunction
\end{algorithmic}
\label{algo:ner}
\end{algorithm}

\begin{figure*}[ht]
    \centering
    \includegraphics[width=\textwidth]{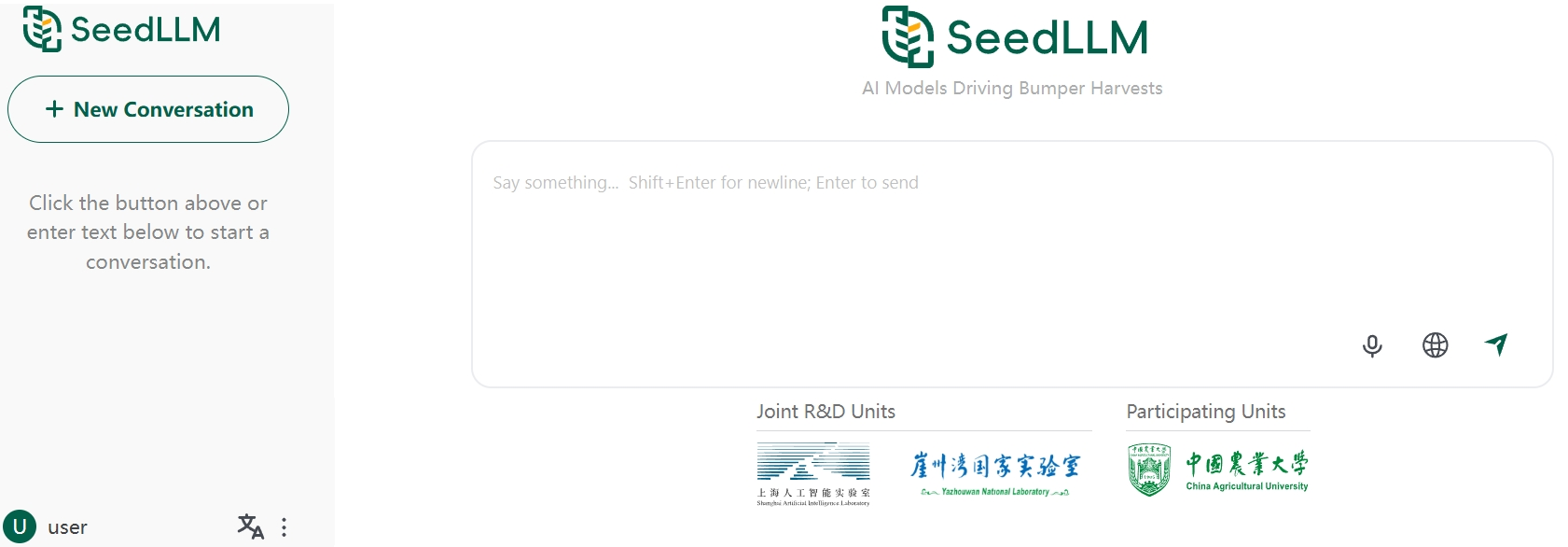}
    \caption{User interface of the deployment platform. The system enables natural language interaction for agricultural knowledge retrieval and question answering.}
    \label{fig:double_column}
\end{figure*}

% \begin{figure*}[ht]
%     \centering
%     \includegraphics[width=\textwidth]{system2.png}
%     \caption{User Interface of ROGRAG. This figure shows the interactive interface of assistant, developed using Gradio. Users can configure pipeline settings, language preferences, and input domain-specific questions for retrieval, with optional multimodal and web search support.}
%     \label{fig:double_column}
% \end{figure*}

\subsection{Retrieval Verifier}
\label{sec:A.4}
Two implementations of iterative retrieval verifier, including  two methods: pre-check and post-check.

\begin{algorithm}[H]
\caption{Retrieval Verifier}
\begin{algorithmic}[1]
    \State Initialize: 
    \\
    \texttt{context} $\gets$ \texttt{logic\_form\_retrieve()}
    \Statex \hspace{\algorithmicindent} % 添加一个缩进，使函数更清晰
    \Function{argument checking}{}
        \If{$LLM\_judge(query, context)$ \\
    $\quad\quad == \text{"support"}$}
    \State \Return $LLM(query, context)$
\EndIf
    \EndFunction
    \Statex \hspace{\algorithmicindent} % 添加一个缩进，使函数更清晰
    \Function{result checking}{}
        \State $reply \gets LLM(query, context)$
        \If{$LLM\_judge(query, context, reply)$\\
    $\quad\quad == \text{"support"}$}
    \State \Return $reply$
\EndIf
    \EndFunction
\end{algorithmic}
\label{algo:check}
\end{algorithm}

% \section{Public Datasets}
% %\section{Data}
% %\subsection{Public datasets}
% \label{sec:B}
% To evaluate the effectiveness of GraphRAG, we utilize two widely adopted benchmark datasets:

% \begin{itemize}[noitemsep, topsep=0pt, leftmargin=*]
%     \item \textbf{UltraDomain}~\citep{qian2024memorag}: A comprehensive dataset capturing domain-specific knowledge across multiple fields.
%     \item \textbf{CMedQA}~\citep{zhang2017chinese}: A specialized dataset for Chinese medical question-answering, mainly used in medical tasks.
% \end{itemize}

% These datasets provide a robust evaluation framework for assessing the retrieval and generation performance of GraphRAG in comparison to alternative methodologies.

\section{Detailed Experimental Setup}
\label{sec:B}
We adopt techniques from \cite{kong2024huixiangdou} regarding the refusal-to-answer task. Specifically, for text splitting, we employ \texttt{ChineseRecursiveTextSplitter}\footnote{\url{https://github.com/chatchat-space/Langchain-Chatchat}} for Chinese text, which takes into account both maximum length and punctuation positions. For English text, we use \texttt{RecursiveCharacterTextSplitter}\footnotemark[5] with a chunk overlap of 32. In both cases, the default chunk size was set to 768 tokens.

\section{Additional Discussion on Experiments}
%\subsection{Experimental discussion}
\label{sec:C}
Our experiments highlight several key factors that affect the performance of the GraphRAG system. Our analysis of iterative NER methods shows that increasing the number of extracted entities can improve accuracy, as erroneous entities will become isolated nodes and will not significantly affect retrieval accuracy. When evaluating the LLM context length, a smaller $rope\_scaling$ yields better performance when the maximum length of the model is sufficient. A larger context length (64k) leads to a slight decrease in accuracy, possibly because the increased volume of information makes it harder to extract meaningful entities and relations. In retrieval and generation analysis, we find that expanding the query representation and incorporating more low-level keys improves accuracy while switching to exact matching leads to performance degradation. This result suggests that fuzzy matching is critical to capturing subtle query semantics, as strict exact matching cannot account for implicit query variations. Comparing retrieval methods, we observe that while the dual-level method achieves higher accuracy, it lacks the ability to provide convincing reasoning on real queries. In contrast, the logic form method provides higher information density and is more concise and clear. Finally, in the validator design, we find that argument checking is more effective than result checking.

% \section{Appendix: User Interface}
% \subsection{User Interface of ROGRAG}

\section{Application}
% % \subsection{Login to Deployment Platform}
% \label{sec:D}
% \begin{itemize}[noitemsep, topsep=0pt, leftmargin=*]
%   \item Online platform: \url{https://seedllm.org.cn}
%   \item Email: wangzhefan@pjlab.org.cn
%   \item Password: www224466
% \end{itemize}

ROGRAG is deployed on
an online research platform SeedLLM\footnote{\url{https://seedllm.org.cn}}, as illustrated in Figure \ref{fig:double_column}. 

% A demonstration video\footnote{\url{https://www.youtube.com/watch?v=NL1_cuIiX-w}} of SeedLLM is provided for reference. 

\end{document}